\documentclass[a4paper,fleqn,usenatbib,onecolumn]{mnras}
\usepackage[T1]{fontenc}
\usepackage{ae,aecompl}
\usepackage{graphicx}	
\usepackage{amsmath}	
\usepackage{amssymb}	
\title[Axisymmetric force-free magnetosphere] 
{Axisymmetric force-free magnetosphere 
in the exterior of a neutron star II:
Maximum storage and open field energies
}
\author[Y. Kojima and S. Okamoto]{
Yasufumi Kojima\thanks{%
E-mail: ykojima-phys@hiroshima-u.ac.jp} and Satoki Okamoto\\
Department of Physics, Hiroshima University, Higashi-Hiroshima, Hiroshima 
739-8526, Japan}
\begin{document}
 \label{firstpage}
 \pagerange{\pageref{firstpage}--\pageref{lastpage}}
 \maketitle
%
\begin{abstract}
A magnetar's magnetosphere gradually evolves
by the injection of energy and helicity from the interior.
Axisymmetric static solutions for a relativistic 
force-free magnetosphere with a power-law 
current model are numerically obtained.
They provide information about the configurations in which the
stored energy is large.
The energy along a sequence of equilibria
increases and becomes sufficient to open the magnetic field.
A magnetic flux rope, in which a large amount of toroidal field is 
confined, is formed in the vicinity of the star, 
for states exceeding the open field energy. 
These states are energetically metastable, and the 
excess energy may be ejected as a magnetar outburst.
\end{abstract}


\section{Introduction}

Solar flares ($E=10^{30}-10^{32}$erg)
are closely related to the Sun's magnetic field.
The flares often give rise to large coronal mass ejections, in which
stored magnetic energy is suddenly converted 
to kinetic energy and radiation.
Giant flares ($E=10^{44}-10^{46}$erg) observed 
in magnetars are widely believed to be analogous, 
but enormously scaled up 
 \citep{2003MNRAS.346..540L,2006MNRAS.367.1594L,2007ApJ...657..967B}.
The flare energy is a part of the total magnetic energy 
($\sim 10^{47} (B_0/10^{14.5} {\rm G})^2 (R/12 {\rm km})^3$).
At smaller energy scales, magnetars also exhibit highly 
variable bursting activity in the X-/gamma-ray band.
This activity and persistent X-ray emission
are powered by the rearrangement and dissipation of
ultra-strong magnetic fields with strengths above $ B_0=10^{14} {\rm G}$
 \citep[e.g.,][for recent review]
 {2015RPPh...78k6901T,2017ARA&A..55..261K}.

The magnetic force is much larger than any other forces in the magnetar 
magnetosphere, and so the force-free approximation may be applicable.
The magnetosphere is twisted by the current flowing in it and the
presence of a toroidal field component is an obvious 
difference from potential magnetic fields in a vacuum. 
The structure changes as the result of the transfer of currents and helicities from 
the interior of the star.
A quasi-steady shearing motion at the base of
a magnetic field twists the exterior field, and the stored energy 
increases at the same time.
When a state exceeds a threshold, the energy is abruptly released 
on a dynamical timescale, leading to energetic flares.
Magnetic energy also builds up as a natural product 
of helicity accumulation. 
The interior itself evolves on a secular timescale 
by Hall drift, ambipolar diffusion, or some other mechanism
 \citep[e.g.,][]{1992ApJ...395..250G,2004MNRAS.347.1273H,2012MNRAS.421.2722K,
 2013MNRAS.434..123V,2014MNRAS.438.1618G,2015PhRvL.114s1101W},
and is affected by the exterior through the boundary.
The interior and exterior are, therefore, coupled to each other.
Recently, \citet{2017MNRAS.472.3914A}
modeled this time-dependent coupled system.
Their evolution model shows that there is no equilibrium solution
for the force-free magnetosphere on timescales of the order 
of thousands of years. This suggests an outburst during that time.
Equilibrium solutions of twisted magnetospheres have been 
considered by a number of authors.
For example, the magnetosphere models for a magnetar
have been numerically constructed as
a part of entire magnetic field structure 
from stellar core to the exterior
\citep{
2014MNRAS.437....2G,2014MNRAS.445.2777F,2015MNRAS.447.2821P,
2017MNRAS.470.2469P}.
 \citet{2016MNRAS.462.1894A}
studied the effect of a covering current-free region on
a twisted magnetosphere. 
The spacetime outside the magnetar is assumed to be flat
in most of these works except for the work of
\citet{2015MNRAS.447.2821P,2017MNRAS.470.2469P}.
Treatment in flat spacetime seems to be reasonable as the
lowest order approximation: a priori, the correction is expected to be not so large,
since the relativistic factor is of order 
$G_{\rm N}M/(Rc^2) \sim 0.2-0.3$ in neutron stars.

In a previous paper \citep{2017MNRAS.468.2011K}, however,
we found that general relativistic effects are significant.
The maximum energies stored in a current-flowing magnetosphere
increases by a factor of a few times from the current-free dipole field energy 
in relativistic models. 
This contrasts with maximum excess energies of 
only a few tens of a percent in non-relativistic models.
This large increase in relativistic models
is related to the formation of a flux rope, 
an axially symmetric torus in the vicinity of the stellar surface, 
when the magnetic field structure is highly twisted.
Curved spacetime helps to confine the torus.
This energy that comes from non-potential magnetic fields 
is available for rapid release through a variety
of mechanisms that may involve instabilities, 
loss of equilibrium, and/or reconnection. 
Is it possible to make a transition from a magnetosphere containing 
a detached magnetic flux to an open field corresponding to mass ejection?
The problem is a dynamical one, but is here examined by comparing 
energies for two different configurations in topology.
One is the energy of equilibrium model, for which 
magnetic field lines are closed and may contain magnetic flux rope.
The other is the energy of open field configuration, for which
all magnetic field lines are open and have the same surface condition
as the equilibrium.
When the energy of a state in a static sequence of models 
exceeds the open field one,
then we may conclude that a transition to a dynamic state must occur. 

This paper is organized as follows. We briefly discuss our model and 
relevant equations for a non-rotating force-free  magnetosphere
in a Schwarzschild spacetime in Section 2.
We then numerically solve the so-called Grad--Shafranov equation
assuming that the current function is given by a 
simple power-law model.
The results are given in Section 3.
Finally, our conclusions are given in Section 4.
We use geometrical units of $c=G_{\rm N}=1$.

\section{Equations}
  \subsection{Magnetic fields}
In this section, we briefly summarize our formalism.
We consider the static magnetic configuration in Schwarzschild space-time 
for the exterior of a non-rotating compact object with a mass $M$. 
The magnetic field for the axially symmetric case is given in terms of 
two functions, 
a magnetic flux function $G$ and a current stream function $S$:
\begin{equation}
\vec{B}=\vec{\nabla}\times \left(
\frac{G}{\varpi}\vec{e}_{\hat{\phi}}
\right)
+\frac{S}{\alpha \varpi}\vec{e}_{\hat{\phi}}
=
\frac{\vec\nabla{G}\times\vec{e}_{\hat{\phi}}}{\varpi}
+\frac{S}{\alpha \varpi}\vec{e}_{\hat{\phi}},
\label{eqnDefBB}
\end{equation}
where $\alpha=(1-2M/r)^{1/2}$ and $ \varpi = r \sin \theta$.
Poloidal current flow is described by 
$4 \pi \alpha \vec{j}_{p}=\vec{\nabla}\times (\alpha \vec{B})$
$=\vec\nabla{S}\times\vec{e}_{\hat{\phi}}/\varpi $.
The components in eq. (\ref{eqnDefBB})
can be explicitly written as
\begin{equation}
[B_{\hat{r}},B_{\hat{\theta}}, B_{\hat{\phi}}]
=\left[\frac{G,_\theta}{r \varpi},
~
 -\frac{\alpha G,_r}{\varpi},
~
 \frac{S}{\alpha \varpi}\right].
\label{Bcomp}
\end{equation}
In the force-free magnetic field, the current function 
$S$ should be a function of $G$, and 
the global structure is determined by
the so-called Grad-Shafranov equation:	
\begin{equation}
\alpha^2 \frac{\partial}{\partial r}
\left( \alpha^2\frac{\partial G}{\partial r} \right)
+\frac{ \alpha^2 \sin\theta}{r^2} \frac{\partial}{\partial \theta}
\left(\frac{1}{\sin \theta}\frac{\partial G}{\partial \theta}\right)
= -\frac{1}{2} \frac{dS^2}{dG} .
\label{eqn:FF}
\end{equation}
In the numerical calculations, we adopt the following model for $S(G)$.
\begin{equation}
 S=\left(\frac{\gamma}{3} \right)^{1/2}G^{3} ,
\label{powerlaw}
\end{equation}
where $\gamma$ is constant, and the
source term in eq.(\ref{eqn:FF}) is simply reduced to $ -\gamma G^5 $
(a power-law model with $n=5$).
This model has been extensively studied in flat spacetime 
for the solar flare model \citep{2004ApJ...606.1210F,2006ApJ...644..575Z,
2012ApJ...755...78Z}.
It is therefore easy to examine relativistic effects. 
It is useful to show a solution in vacuum for eq.(\ref{eqn:FF}) with $S=0$.
The magnetic function $G$ is expanded in terms of Legendre polynomials $P_{l}(\theta)$:
\begin{equation}
 G(r,\theta )=-\sum_{l \ge 1}g_{l}(r)\sin \theta 
\frac{d P_{l}(\theta )}{d\theta },
   \label{Sexpd.eqn}
\end{equation}
and the radial functions $g_{l}$ 
are given by an analytic function.
For example, the radial function for a dipole ($l=1$) is
\begin{eqnarray}
\label{eqnpuredip}
g_{1} &=& - \frac{3 B_{0} R^3 r^2}{8 M^3}\left[
\ln\left(1-\frac{2M}{r}\right) 
+\frac{2M}{r}+\frac{2M^2}{r^2} \right]
\\
&\approx &
\frac{ B_{0} R^3 }{r}\left[
1+\frac{3M}{2r}+\frac{12M^2}{5r^2} +\cdots \right],
~~~~~ (M/r \ll 1)
\nonumber
\end{eqnarray}
where $ B_0 $ is the typical field strength and 
$R$ is the stellar surface radius.
In this expression, the first expression is an exact solution
and the second is its approximation in the weak gravity regime
$M/r \ll 1$.
The magnetic dipole moment $\mu$ is given by $ \mu = B_0 R^{3}$
from the asymptotic form of $g_{1}$ at infinity.
The field strength at the surface pole is $2B_{0}$ in a non-relativistic model, 
whereas it is larger by a factor of order 
${\mathcal O}(M/R)$ in a relativistic model with the same dipole moment $\mu$.
In this paper, we use as the normalization factor the field $B_{0}$,
which is defined by the dipole moment $\mu$, but does not denote 
the field strength $2B_{\hat r}(R,0)$
at the surface pole except for the case $M/R=0$.

  We now discuss the boundary conditions needed to
solve eq.~(\ref{eqn:FF}) with (\ref{powerlaw}).
Along the polar axis, the magnetic function $ G$ should satisfy the 
regularity condition $ G=0$ at $\theta =0$ and $ \pi$.
Asymptotically ($r \to \infty $),
the function should decrease as $G \propto r^{-1} $. 
At the stellar surface $r=R$, 
the magnetic function $G$ is assumed to be a dipolar ($l=1$) field:
\begin{equation}
 G_{S}(\theta ) \equiv G(R,\theta )=g_{1}(R)\sin^2 \theta , 
   \label{SurfG.eqn}
\end{equation}
where $ g_{1}(R)$ is given by eq. (\ref{eqnpuredip}).
The numerical method for solving the non-linear equation
 (\ref{eqn:FF}) with (\ref{powerlaw}) is described in 
 \cite{2017MNRAS.468.2011K}.

  \subsection{Helicity and energy}
\begin{figure}
\centering
\includegraphics[scale=1.0]{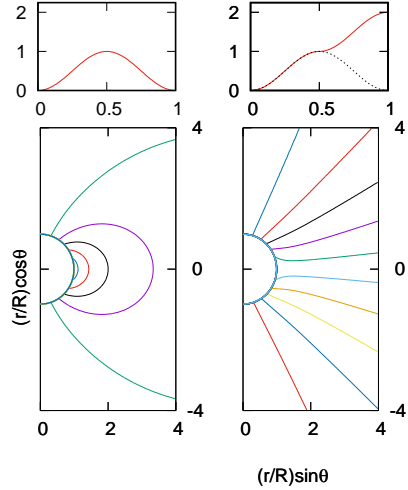}
\caption{
Magnetic field lines for a potential dipole (left panel) and 
an open field (right panel).
In the top panels, the magnetic functions 
$G_{S}(\theta)/g_{1}(R)$ and $ G^{*} _{S}(\theta) /g_{1}(R)$ at the surface
are shown by a solid line.
All dipolar field lines are closed in the left panel, 
whereas those of the open field extend from the surface to infinity.
The radial component $B_{\hat r} $
is positive on the surface in the right panel.
A realistic field is obtained by changing
the sign of $B_{\hat r} $ in the southern hemisphere, and
locating the current sheet for the split-monopole-like configuration 
on the equator, 
although this is not necessary for this work.
}
\label{fig:1}
\end{figure}

  Two integrals,  magnetic helicity and energy, are useful to 
characterize equilibrium solutions of the magnetospheres.
Magnetic helicity represents a global property of magnetic fields,
and is obtained by integrating the product of
two vectors, namely, ${\vec A}$ and
 $ {\vec B} (= {\vec \nabla} \times {\vec A})$.
A gauge-invariant quantity, $H_{\rm R} $, is defined by 
the difference of the magnetic helicity of a force-free field
from that of a potential field with the same surface boundary condition.
The total relative helicity in the exterior $(r \ge R)$ 
is given by 
\begin{equation}
H_{\rm R} 
= \int_{ r\ge R} {\vec A}\cdot {\vec B} \sqrt{g_3}d^3x
= 4\pi\int_{ r\ge R}  \frac{ GS}{\alpha^{2}}
\frac{dr d\theta }{\sin \theta} ,
\label{eqnHR}
\end{equation}
where $ \sqrt{g_{3}} (= \alpha^{-1} r^2 \sin \theta)$ is the 
determinant of the 3-dimensional space metric
\citep{2017MNRAS.468.2011K}.
Magnetic energy stored in the force-free magnetosphere is also 
given by integrating over a 3-dimensional volume: 
\begin{equation}
E_{\rm EM} = \int_{ r\ge R} \frac{\alpha B^2}{8\pi} \sqrt{g_3}d^3x
=\frac{1}{4} \int _{ r\ge R} B^2 r^2 \sin \theta dr d\theta 
= \frac{1}{4}\int_{ r\ge R} \left[ \left( \alpha  
\frac{\partial G}{\partial r} \right)^2
+ \left( \frac{1}{r}\frac{\partial G}{\partial \theta} \right)^2
+\left(\frac{S}{\alpha} 
\right)^2  \right] 
\frac{dr d\theta }{\sin \theta} .
\label{eqn:Eng}
\end{equation}
In eq. (\ref{eqn:Eng}), the factor $\alpha$ in front of
$B^2$ may be understood by considering the Maxwell equations
in curved spacetime. Equivalently, 
the expression (\ref{eqn:Eng}) can also be obtained 
by $\int T^{t} _{t}  \sqrt{-g_{4}} d^3x$
in terms of the energy momentum tensor $T^{t} _{t} $ and 
the determinant of the 4-dimensional spacetime metric $\sqrt{-g_{4}}$
\citep{2017MNRAS.468.2011K}.
 The numerical results for $H_{\rm R} $ and $E_{\rm EM}$
in a force-free magnetosphere, 
which depend on the twist, will be given in the next section.

  We here discuss the energy for two reference configurations.
For a given dipolar field at the surface (\ref{SurfG.eqn}), 
the lowest energy state is given by the potential field.
This energy is denoted by $E_{0}$, and is given by
$ E_{0}= B_{0}^2R^3/3$ for a dipolar potential field in flat spacetime
\citep[e.g.,][]{1993ApJ...410..412L}.
The value increases in relativistic models; for example, it has been 
numerically calculated as $ E_{0} = 0.74 B_{0}^2R^3$ 
for a model with $M/R=0.25$
\citep{2017MNRAS.468.2011K}.

 Another important criterion is the open field energy $ E_{\rm open}$.
The open field configuration is demonstrated in Fig. \ref{fig:1}.
Suppose that initially closed magnetic field
lines of a force-free magnetosphere
are stretched out to infinity by some artificial means, 
keeping the same boundary condition.
Additional energy is necessary to open it.
When the energy $E_{\rm EM} $ of a force-free magnetosphere 
is less than $ E_{\rm open}$, opening is difficult.
When  $E_{\rm EM} $ > $ E_{\rm open}$,
an open field configuration is energetically preferable.
An abrupt transition to the open field 
may be related to the mass ejection in flares. 
It is therefore important to examine 
whether or not there exists a state with $E_{\rm EM} >  E_{\rm open}$.
  The calculation of $E_{\rm open}$ has been discussed previously 
\cite[e.g.,][]{1993ApJ...410..412L}.
Here, we briefly summarize the procedure.
We modify the boundary condition (\ref{SurfG.eqn}) at the surface as
\begin{eqnarray}
G^{*} _{S}(\theta) &= & G_{S}(\theta) 
~~~~~~~~~~~~~~( 0 \le \theta \le \pi/2 ),
\nonumber
\\
G^{*} _{S}(\theta) 
&=& 2g_{1}(R) -G_{S}(\theta)
~~( \pi/2 < \theta \le \pi ).
\label{eqn:flip}
\end{eqnarray}
Note that $G_{S}(\pi/2) =g_{1}(R)$ and $G_{S}(\theta)$
is a continuous function on the whole range
$0 \le \theta \le \pi$.
The functions $G_{S}(\theta)$ and $ G^{*} _{S}(\theta) $
are displayed in the top panels of Fig.\ref{fig:1}.
By solving eq.(\ref{eqn:FF}) with $S=0$ and 
surface boundary condition $G_{S}(\theta)$,
we have a dipolar potential field as shown 
in the left panel of Fig.\ref{fig:1}.
By replacing the boundary condition with $ G^{*} _{S}(\theta) $,
an open field solution $G_{\rm open}$ is obtained,
as shown in the right panel of Fig.\ref{fig:1}.
The boundary condition $ G^{*} _{S}(\theta) $ is 
a monopolar magnetic field; that is,
radial component $B_{\hat r}= |B_{\hat r}(\theta)|$ 
is one-way direction at the surface, 
so that all the field lines extend to infinity.
The desired solution is obtained by taking this unphysical magnetic field 
and reversing its direction only on those lines in the southern hemisphere 
($\pi/2 < \theta \le \pi$).
The magnetic energy is unchanged by this sign-flipping,
and may be calculated for the solution $G_{\rm open}$.
The result is $ E_{{\rm open}}/E_{0} =1.66$ 
for a dipolar field in flat spacetime
\citep[e.g.,][]{1993ApJ...410..412L,2004ApJ...606.1210F}.
The open field is strict poloidal, with $B_\phi = 0$,
although the force-free field is twisted with $B_\phi\ne 0$.
A finite twist is assumed to propagate to infinity along open field lines. 
 The field necessarily includes a current sheet on 
the equator, which separates the regions of opposite magnetic polarity. 
It is instructive to approximate magnetic function $G_{\rm open}(r, \theta)$
as a monopole solution $G_{M}(r, \theta)\equiv g_{1}(R) (1-\cos\theta)$.
The function $G^{*} _{S}(\theta) $ is very close to 
$G_{M}(R, \theta)$ at the surface, but has some
higher multi-poles with small amplitudes.
The magnetic energy $E_{{\rm monopole}}$ is calculated
as $ g_{1}(R) ^2/(2R) $, which is reduced to
$E_{{\rm monopole }}/E_{0} =1.5$  in flat spacetime.
The open field energy  $ E_{{\rm open}}$ contains
16 \% contribution from higher multi-poles.
%

  \subsection{Virial}
 
Here we derive some useful relations concerning total magnetic energy.
We multiply eq. (\ref{eqn:FF}) by $F \partial G/\partial r$, 
where $F$ is an arbitrary function of $r$, and integrate over the  space  outside a radius $R$.
Using integration by parts, we have the identity:
\begin{eqnarray}
&&\frac{1}{4} \int _{R}^{\infty} \int _{0}^{\pi}
\alpha^{2} r^2 \frac{d F}{d r} 
\left(B_{\hat r}^2+B_{\hat \theta}^{2}+B_{\hat \phi}^{2}\right)
dr \sin\theta d\theta
\nonumber
\\
&&=\frac{1}{4}\int _{0}^{\pi}
\left[ \alpha^{2} r^{2} F(B_{\hat r}^2-B_{\hat \theta}^{2}-B_{\hat \phi}^{2})
  \right] _{r=R}
 \sin\theta d\theta
+\frac{1}{4}\int _{R}^{\infty} \int _{0}^{\pi}
\frac{r^{4}}{F} \frac{d }{d r} \left( \frac{\alpha^2 F^2}{r^2}\right)
B_{\hat r}^{2} dr \sin\theta d\theta ,
\label{eqnvir}
\end{eqnarray}
where we have used the components (\ref{Bcomp}) of magnetic fields, and 
assumed that ${\vec B} $ approaches zero at infinity. 
As a first application, 
we consider this formula in flat spacetime by setting $\alpha =1$.  
By choosing $F=r$, the left hand side in eq.(\ref{eqnvir}) is reduced to
the magnetic energy $E_{\rm EM}$ stored in the exterior $r \ge R$,
and the volume integral part of the right hand side vanishes.
Thus, the magnetic energy is expressed by the surface term, that is, the virial theorem 
\citep{1961hhs..book.....C, 2004ApJ...606.1210F}:
%
\begin{equation}
E_{\rm EM} =\frac{1}{4}  \int _{0}^{\pi}
\left[r^3(B_{\hat r}^2-B_{\hat \theta}^{2}-B_{\hat \phi}^{2})\right] _{r=R}
\sin\theta d\theta.
\label{eqnvir0}
\end{equation}
Since $ E_{\rm EM}\ge 0$, we have an inequality for magnetic components at $r=R$:
\begin{equation}
\int _{0}^{\pi} \left[ B_{\hat r}^2 \right] _{r=R}\sin\theta d\theta
\ge 
\int _{0}^{\pi} \left[B_{\hat \theta}^{2} +B_{\hat \phi}^{2}\right]_{r=R}
 \sin\theta d\theta
\ge
\int _{0}^{\pi} \left[B_{\hat \phi}^{2} \right]_{r=R}\sin\theta d\theta.
\label{eqnbrbt}
\end{equation}
We consider a sequence of solutions with fixed boundary condition 
 (\ref{SurfG.eqn}), which means that the radial component 
($B_{\hat r} \propto G_{,\theta}$) is always fixed at the surface.
Equation (\ref{eqnbrbt}) constrains the toroidal component 
$B_{\hat \phi} \propto \gamma ^{1/2}$.
Thus, there is a maximum of $\gamma$ 
\citep{2004ApJ...606.1210F}.
%

Extension to the relativistic case with $\alpha \neq 1$
needs a little care, since the left hand side with $F=r$
in eq.(\ref{eqnvir}) is no longer $E_{\rm EM}$. It
differs by a factor $\alpha^2$ (See eq.(\ref{eqn:Eng})).
Some calculations provide
\begin{equation}
 E_{\rm EM}=\frac{1}{4} 
\int _{0}^{\pi}\left[\alpha^2 r^3( 
B_{\hat r}^2-B_{\hat \theta}^{2}-B_{\hat \phi}^{2})\right]_{r=R}\sin\theta d\theta
+\frac{1}{4} \int _{R}^{\infty} \int_{0}^{\pi}
r^{2}(1-\alpha^{2})(2B_{\hat r}^2+B_{\hat \theta}^2+B_{\hat \phi}^{2})
dr \sin\theta d\theta .
\label{eqnvr1}
\end{equation}
This was derived in \cite{2011ApJ...738...75Y}.
Here the volume integral is included in the expression for $E_{\rm EM}$.
Another expression for $E_{\rm EM}$ is also possible.
By choosing a tortoise coordinate 
$F=r_{*}$$(\equiv r+ 2M\ln(r/2M-1))$, 
which satisfies $d  r_{*} /d r =\alpha^{-2}$, 
the left hand side in eq.(\ref{eqnvir}) 
is reduced to  $E_{\rm EM}$, and is the equation can be written as 
\begin{equation}
E_{\rm EM}=\frac{1}{4}
\int _{0}^{\pi}\left[\alpha^2 r_* r^2( 
 B_{\hat r}^2-B_{\hat \theta}^{2}-B_{\hat \phi}^{2}) \right]_{r=R}
\sin\theta d\theta
+\frac{1}{2}\int _{R}^{\infty}\int _{0}^{\pi}
(r^2-r r_{*}+3Mr_{*})B_{\hat r}^2
\sin\theta drd\theta .
\label{eqnvr2}
\end{equation}
The relativistic expressions (\ref{eqnvr1}) and (\ref{eqnvr2}) 
represent the fact that the amount of exterior magnetic energy is determined 
not only by the surface values but also by some volume integral, 
unlike in the non-relativistic case (\ref{eqnvir0}).
The additional term is positive in reasonable
stellar models, and the radial component $B_{\hat r}^2$ is dominant there.
The term is of order $E_{\rm EM}\times (M/R)$, and
acts to nonlinearly increase $E_{\rm EM}$,
when $M/R$ is not very small.
That is, a correction of $E_{\rm EM}$ further increases $E_{\rm EM}$ itself. 
Thus, a state having large $E_{\rm EM}$
is less sensitive to the surface boundary in a relativistic system.
These expressions explain the properties of an interesting structure,
a soliton-like  magnetic flux rope, found in the numerical models.

Finally, if we choose $F= r/\alpha$, then
the surface integral is given by a volume integral as
\begin{equation}
\frac{1}{4} \int _{0}^{\pi}\left[\alpha^2 r^3( 
B_{\hat r}^2-B_{\hat \theta}^{2}-B_{\hat \phi}^{2})\right]_{r=R}
\sin\theta d\theta =
\frac{1}{4} \int _{R}^{\infty}\int _{0}^{\pi}
 \alpha^{-1}r(r-3M)
(B_{\hat r}^2+B_{\hat \theta}^2+B_{\hat \phi}^{2})
dr \sin\theta d\theta . 
\end{equation}
As long as $R> 3M$,
the right hand side is positive definite, so that we have
the same relation (\ref{eqnbrbt})
as in flat spacetime. 

\section{Numerical results}

  A sequence of magnetospheres is numerically constructed
for a fixed boundary condition (\ref{SurfG.eqn}) at the surface.
We start with a potential field solution, and follow the change of structure 
by increasing the toroidal magnetic field
for a fixed relativistic factor $M/R$.
A simple method is increasing the parameter 
$\gamma$ in eq.(\ref{powerlaw}).
Numerical solutions are however limited 
by the method as discussed below:
the higher energy branch of solutions cannot be obtained
An alternative method, which is used for the same power law
current model in flat spacetime\citep{2004ApJ...606.1210F,
2006ApJ...644..575Z,2012ApJ...755...78Z},
is increasing azimuthal flux or helicity as the degree of twist.
The constant $\gamma$ is determined a posteriori.
Thus, both magnetic energy $E_{\rm EM}$ and relative helicity $H_{\rm R}$
are a multi-valued function of $\gamma$.
A similar method is used in a different model
\citep{2015MNRAS.447.2821P,2018MNRAS.474..625A}, where the
physical extent of a field line is specified first, and
the corresponding toroidal field strength is determined as the result.

\begin{figure}
\centering
\includegraphics[scale=0.75]{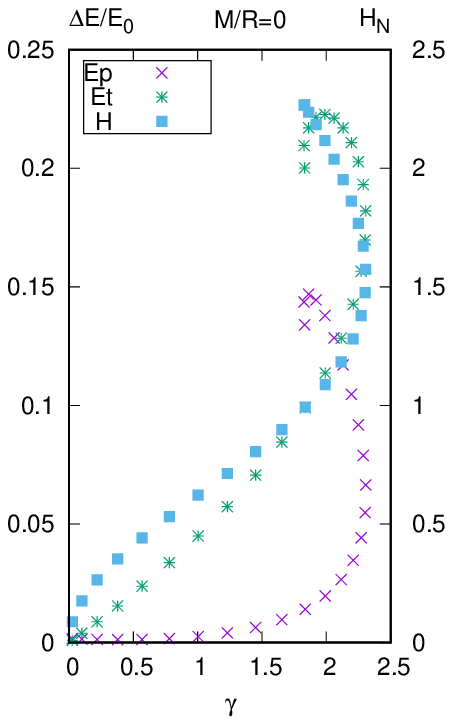}
\includegraphics[scale=0.75]{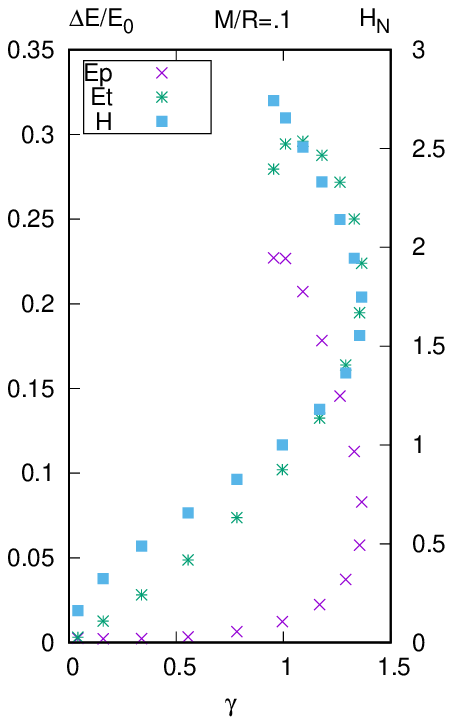}
\includegraphics[scale=0.75]{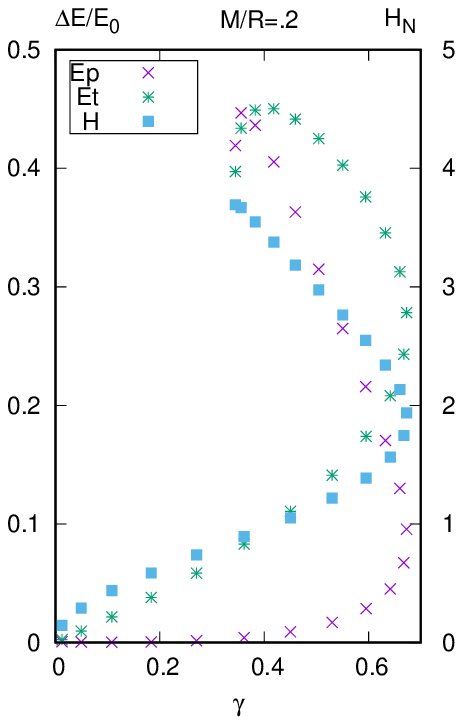}
\includegraphics[scale=0.75]{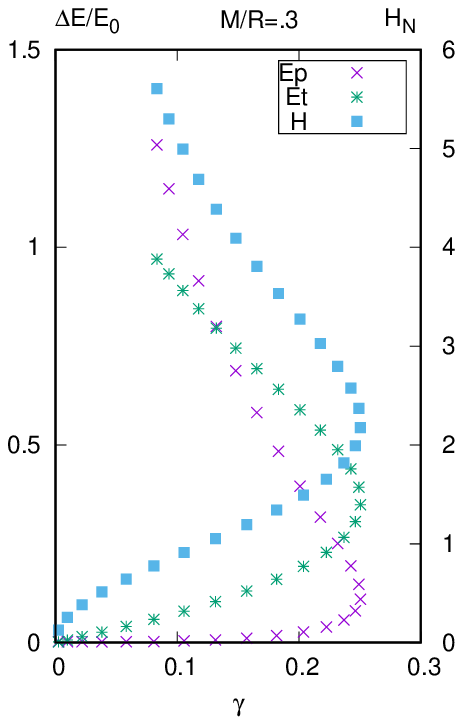}
\caption{Increase in magnetic energy $\Delta E/E_0$ 
from the potential dipole field is shown in the left scale and
total relative helicity $H_{N}=H_{\rm R} /(4\pi E_0 R)$
is shown in the right scale.
The energy of the poloidal component is
denoted by crosses, that of the toroidal component by asterisks,
and helicity by squares.
The horizontal axis denotes 
the dimensionless value $\gamma(B_0 R^2)^5 $.
From left to right, the panels show the results for 
$M/R=0, 0.1, 0.2, 0.3$.
}
\label{fig:2}
\end{figure}

   The magnetic energy and the relative helicity for the models 
with $M/R =0, 0.1, 0.2, 0.3$ are shown in Fig. \ref{fig:2}. 
For a better understanding of the mechanism, the energy 
difference $\Delta E(=E_{\rm EM}-E_{0})$ is divided as
$\Delta E =\Delta E_{t}+\Delta E_{p}$,
into a toroidal component and a poloidal component.
The general tendency is the same in all models.
There is a maximum of $\gamma$, and there are two branches 
in the curves of $\Delta E_{p}$,  $\Delta E_{t}$ and $H_{\rm R}$, 
when we consider solutions as a function of $\gamma$. 
In the lower branch, an increase of $\Delta E_{t} $ is evident,  
whereas $\Delta E_{p}$ is almost zero.
The toroidal energy $\Delta E_{t} $ monotonically increases with $\gamma$,
since $B_{\hat \phi} ^2 \propto \gamma $ by eq.(\ref{powerlaw}). 
However, there is a certain limit to $B_{\hat \phi} ^2$ or
$ \gamma$ by eq.(\ref{eqnbrbt}). 
After passing the turning point of $\gamma$, 
$\Delta E_{p}$  increases dramatically in the upper branch.
This means that the poloidal field structure significantly  
changes from that of the potential field
in order for a larger toroidal field to be supported. 
The curve of $\Delta E_{p}$ or $\Delta E_{t}$
in Fig. \ref{fig:2} no longer goes up,
but curls into a limiting point with a further increase of twist.
This behavior is similar to that often appearing
near a critical point in nonlinear dynamics.
\citet{2004ApJ...606.1210F,2006ApJ...644..575Z,2012ApJ...755...78Z}
have shown the detailed behavior of this current model
in a flat spacetime. A careful treatment is necessary when changing
the parameter near the endpoint.
In this work, we do not resolve the endpoint of the sequence, 
because the maximum value of energy or helicity
is unchanged even if we approach the termination more closely.
The maximum of the ratios $\Delta E_{p}/E_{0}$ and
$\Delta E_{t}/E_{0}$ increase with $M/R$.
There is a qualitative difference between 
the model with $M/R=0$ and that with $M/R=0.3$.
At the maximum,  we have 
$\Delta E_{t} > \Delta E_{p}$ in the former,  while 
$\Delta E_{t} < \Delta E_{p}$ in the latter.
Near the endpoint, the ratio is
 $E_{t}/E_{p}=\Delta E_{t} /(E_{0}+\Delta E_{p}) <  $ 0.2-0.4,
that is, the poloidal field is always dominated for
stable configurations.
%

\begin{figure}
\centering
\includegraphics[scale=1.2]{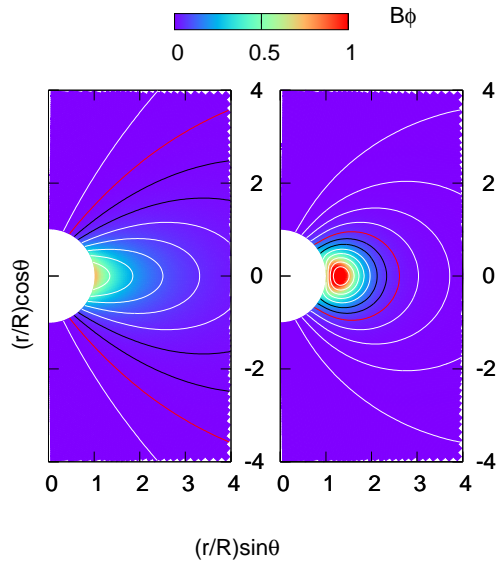}
\caption{Contour lines of the magnetic function $G$
and color contours of normalized $B_{\hat{\phi}}(\ge 0)$ 
in the $r$-$\theta$ plane.
The structure corresponds to the model of maximum energy 
in for the sequences with $M/R=0$ (left panel)
and $M/R=0.3$ (right panel).
 }
 \label{fig:3}
 \end{figure}

Figure \ref{fig:3} shows the magnetic field structure 
of a highly twisted state, 
i.e,  near the endpoint along a sequence for each model, 
at which the stored energy is a maximum.
We compare the model for $M/R=0$ with that for $M/R=0.3$. 
In the figure, we show the magnetic function $G$ by contour lines,
and the toroidal component $B_{\hat \phi}$ in the 
$r$-$\theta$ plane by colors.
Only the interior part is shown, since the field outside
approaches a vacuum solution due to $S \propto G^3 \to 0$, and
so the outer part does not change.
Magnetic lines are stretched toward the exterior by a strong twist.
The maximum of the toroidal magnetic field $B_{\hat \phi}$
is located near the surface for the model with $M/R=0$.
The topology of the magnetic function for the model
$M/R=0.3$ is different. 
There are loops of field lines 
around the center $(r/R, \theta) =(1.2, \pi/2)$,
and the maximum of $B_{\hat \phi}$ occurs there.
The structure represents a flux rope braided by toroidal and poloidal 
magnetic fields in three-dimensional space.
The magnetic flux is likely to expand, but
general relativistic effects suppress the expansion,
and allow a larger amount of magnetic energy to be stored
at the same time.
It should be noted that the similar flux-rope structure was also found 
in previous results in literature.
For example, \citet{2015MNRAS.447.2821P} obtained it
in the exterior model of a neutron star, by using
a different current model and numerical method.
In their numerical method, the radial extent of current-flowing field lines 
is specified to calculate a static solution.
When the region extends to several times the stellar radius, then
a remarkable flux-rope can be seen.
\citet{2018MNRAS.474..625A} applied the similar method  
to a model in flat space-time, and obtained it.
A direct comparison is difficult due to the differences in both models. 
However, by comparing our models with different relativistic factor,
the flux-rope formation is not inherent in general relativity,
but is sustained by the effect.
%

\begin{figure}
\begin{minipage}{0.45 \hsize}
\centering
\includegraphics[scale=0.8]{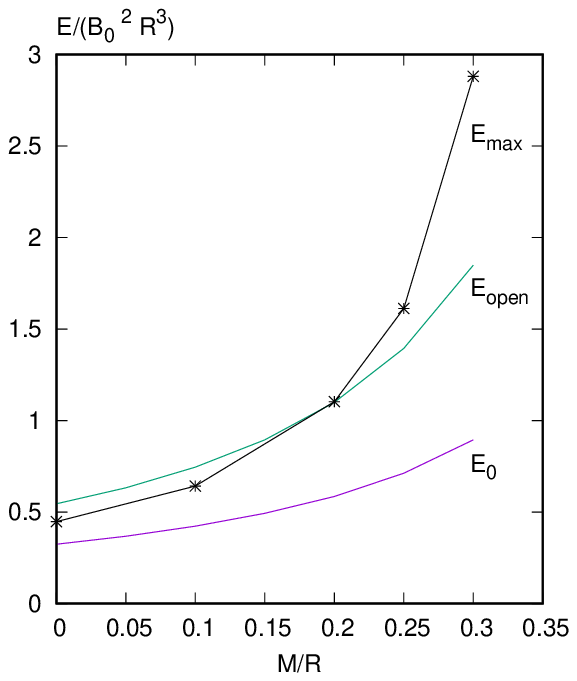}
\caption{ Maximum energy $E_{\rm max}$, denoted by asterisks,
as a function of the relativistic factor $M/R$.
The potential field energy $E_0$ and open field energy $E_{\rm open}$
are also displayed. These energies are normalized by $B_{0} ^2 R^3$.
}
\label{fig:4}
\end{minipage}
\hspace{10mm}
\begin{minipage}{0.45 \hsize}
\centering
\includegraphics[scale=1.2]{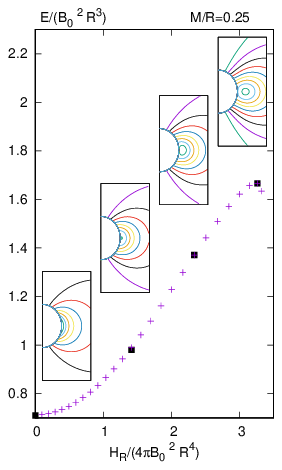}
\caption{ Total magnetic energy $E_{\rm EM}/(B_{0} ^2 R^3)$ is shown 
as a function of total relative helicity 
$H_{\rm R} /(4\pi B_{0} ^2 R^3) \approx 0.71 H_{N}$
along a sequence of models with $M/R=0.25$.
Typical magnetic field configurations are also displayed in the figure.
They correspond, from the bottom to the top, to a potential dipole, the maximum of $\gamma$,
the open field energy $E_{\rm open}$ and the maximum energy.
Their energies are indicated by squares. 
}
\label{fig:5}
\end{minipage}
\end{figure}

 Figure \ref{fig:4} shows three energies, $E_0$, $ E_{\rm max}$
and $E_{\rm open}$ normalized by $B_{0}^2 R^3(=\mu^2 /R^3)$,
as a function of the relativistic factor $M/R$.
The potential field energy $E_0$ is the minimum, and the maximum 
$ E_{\rm max}$ is calculated along a sequence of force-free magnetosphere models.
As inferred from eq.(\ref{eqnpuredip}), 
$E_{0}~ (\propto g_{1}(R)^2)$ increases with $M/R$
in our normalization for fixed magnetic dipole moment.
Figure \ref{fig:4} shows the open field energy $E_{\rm open}$ 
also increases with $M/R$. 
The ratio $E_{\rm open}/E_{0}$  depends less on
the normalization, but it also slightly increases.
For example, $E_{\rm open}/E_{0} =1.66$ at $M/R=0$
\citep{1993ApJ...410..412L,2004ApJ...606.1210F}
and it increases to 2.05 at $M/R=0.3$.
The increase of $ (E_{\rm open} -E_{0})/E_{0}$
means a large load energy is required to open field, and 
it seems to be more difficult to
make the transition from closed to open configurations
in more relativistic system.
However, the curve of maximum energy $ E_{\rm max}$ 
with $M/R$ is steeper, as shown in Fig. \ref{fig:4}.
Thus, a state with $E_{\rm EM} >E_{\rm open}$ 
is realized in a relativistic system with $M/R > 0.2$.
The steep increase
is closely related to the formation of a detached flux rope.
The excess energy $E_{\rm EM} -E_{\rm open}$ is
released with the flux rope eruption.
The maximum energy is, for example,
$(E_{\rm max}-E_{\rm open} )/E_{0} $$=0.33$ for the model with
$M/R=0.25$ and 1.19 for the model with $M/R=0.3$.
%

Figure \ref{fig:5} shows total magnetic energy along 
an increasing sequence of relative helicity for a model
with $M/R =0.25$.
The magnetic function is also shown by contours for 
four representative states. They are characterized, in increasing energy, by
a potential field, the maximum of $\gamma$, an energy equal
to the open field case $E_{\rm EM}=E_{\rm open}$ and  
the maximum energy $E_{\rm max}$. A flux rope is evident after passing the turnover of $\gamma$
(i.e., in the upper branch in Fig. \ref{fig:2}).

\section{Conclusion}

  In this paper, we have studied energy storage
in a relativistic force-free magnetosphere with power-law current model.
Total magnetic energy increases as the helicity increases 
in axially symmetric equilibria.
An evolution scenario of twisting magnetospheres is constructed
through a quasi-static sequence of equilibrium states.
That is, the magnetosphere over a long timescale gradually
changes so as to accumulate magnetic helicity.
The helicity stored in magnetosphere
decreases only in a dynamical process.
In general, total helicity is conserved
as far as the ideal MHD condition ${\vec E} \cdot {\vec B} = 0 $
holds. 
A catastrophic change, for 
which the acceleration field  ${\vec E} \cdot {\vec B} \ne 0 $ 
should be relevant, may be an outburst.
It is interesting to note that larger energy 
and helicity are capable of being stored in a relativistic 
magnetosphere than in a non-relativistic one. 
The energy at the endpoint 
along our equilibrium sequence with $M/R \ge 0.2$
exceeds the open field energy.
This means that the high-energy states with $E_{\rm EM} > E_{\rm open}$ 
are metastable.
A transition to a lower energy state is associated with the eruption 
of a magnetic flux rope.
This is observed as a magnetar flare.
It is, however, not clear at the moment how much energy is ejected.
It depends on the stability of the high-energy states.
That is, the excess $E_{\rm EM} - E_{\rm open}$ is
almost zero when instability sets in soon after reaching a state with
$E_{\rm EM} = E_{\rm open}$.
On the other hand, the amount of energy increases,
when the high-energy state is more stable 
and energy is built up before a bursting event.
Such a problem requires a dynamical method for its solution
\citep[e.g.,][as resistive simulation in flat spacetime]
{2012ApJ...746...60L,2013ApJ...774...92P,2014PTEP.2014b3E01K},
which is beyond the scope of the quasi-equilibrium approach used here. 
The present paper as well as similar studies
\citep[e.g.,][]{2004ApJ...606.1210F,2007ApJ...660.1683W,
2016MNRAS.462.1894A}
are useful to explore and describe conditions that result 
in equilibrium solutions containing substantial energy. 
By combining these works, it is evident that
a large amount of the energy stored is related with flux rope formation
in the vicinity of the surface.
The maximum energy of detached configuration exceeds the open field energy,
so that a transition to the lower energy state is possible. 
In the dynamical transition, a flux rope may be ejected.
There are at least three elements studied so far that increase the energy 
stored in a force-free magnetosphere.
Two are related to the current model, 
so that we, for convenience, assume the power law form
$(B_{\phi} ^2 \propto) S^2 \propto G^{n+1}$ (see eq. (\ref{powerlaw}).)
%
As the power index $n$ increases, 
the distribution of the toroidal magnetic field becomes steeper. 
The flux rope is formed due to strong confinement 
and the maximum energy increases.
When the index $n$ is larger than 9,
the energy exceeds the dipolar open-field energy by a few percent
\citep{2004ApJ...606.1210F}.
The second important element is covering by an external current-free 
magnetic field. 
The model can be described as
$S^2 \propto (G-G_{c})^{n+1}$ for $G \ge G_{c} $,
while $S=0 $  for $G<G_{c} $.
The cut-off means that current flowing is spatially limited.
The interior non-potential field is held down, and energy storage is 
enhanced. For example, \citet{2007ApJ...660.1683W} found that 
the maximum excess energy is 18\% of the 
dipolar potential field energy.
The third element is confinement by curved space-time, 
considered here and in a previous paper.
General relativistic effects suppress the outward eruption of magnetic flux,
and relativistic models are
capable of storing significantly more energy than the 
corresponding potential energy.
The excess is 30\% for $ M/R= 0.25$ and 119 \% for $ M/R=0.3$.

The maximum of the buildup energy also depends on 
other factors, such as the magnetic field at the surface
\citep{2012ApJ...750...25W}.
At moment, it is not clear which factors are important, since we do not 
know the correct current model and surface condition
of a magnetar.
However, it is a relativistic object, so general relativistic 
effects should be taken into account in any model.

\section*{Acknowledgements}
This work was supported by JSPS KAKENHI Grant Numbers 
JP26400276 and JP17H06361.

 \bibliographystyle{mnras}
 \bibliography{kojima18jan} 

\end{document}